\RequirePackage{etex}

\documentclass{aa}  
\usepackage{graphicx}
\usepackage{txfonts}
\usepackage{float}
\usepackage[dvipsnames]{xcolor} 

\usepackage{hyperref}
\hypersetup{
    colorlinks,
    linkcolor=NavyBlue,
    filecolor=NavyBlue,    
    urlcolor=NavyBlue,
    citecolor=NavyBlue
    }

\newcommand{\HRIEUV}{HRI$_{\mathrm{EUV}}~$}
\newcommand{\HRIEUVNOSPACE}{HRI$_{\mathrm{EUV}}$}

\def\code#1{\texttt{#1}}

\begin{document}

    \title{Small-Scale and Transient EUV Kernels in Solar Flare Ribbons}

   \author{Hannah Collier
          \inst{1,2}
          \and
          Säm Krucker \inst{1,3}\
        \and
        Laura A. Hayes \inst{4}\
        \and
        Emil Kraaikamp \inst{5}\
        \and
        David Berghmans \inst{5}
        \and
        Daniel F. Ryan \inst{6}
          }

   \institute{University of Applied Sciences and Arts Northwestern Switzerland (FHNW), Bahnhofstrasse 6, 5210 Windisch, Switzerland\\          \email{hannah.collier@fhnw.ch}
         \and
         ETH Z\"{u}rich,
        R\"{a}mistrasse 101, 8092 Z\"{u}rich, Switzerland
        \and
         Space Sciences Laboratory, University of California, 7 Gauss Way, 94720 Berkeley, USA
         \and
        Astronomy \& Astrophysics Section, School of Cosmic Physics, Dublin Institute for Advanced Studies, DIAS Dunsink Observatory, Dublin, D15 XR2R, Ireland
         \and
          Solar-Terrestrial Centre of Excellence (SIDC), Royal Observatory of Belgium, Ringlaan 3 Av. Circulaire, 1180 Brussels, Belgium
          \and 
          University College London, Mullard Space Science Laboratory, Holmbury St Mary, Dorking, Surrey RH5 6NT UK
             }

   \date{Received December 16th, 2025; accepted March 12th, 2026}

  \abstract
   {}
   {Flare ribbons form when energy released by coronal magnetic reconnection is deposited in the low layers of the solar atmosphere. Therefore, by studying the dynamics of flare ribbons, one obtains an indirect measurement of reconnection in the corona. The aim of this work is to quantify the spatial and temporal scales of substructures within the Extreme Ultraviolet (EUV) flare ribbons, known as kernels, as a probe of the spatial extent and duration of energy injection during the impulsive phase of solar flares.}
   {To do this, unprecedented observations of an M2.5 GOES-class flare from the March 2024 major flare campaign of Solar Orbiter were used. These data were obtained at high-cadence in short-exposure mode with the Extreme Ultraviolet Imager's high-resolution telescope, \HRIEUVNOSPACE. Individual kernels were automatically identified using a classical computer vision algorithm. From this, size distributions of ribbon kernels were derived, and an average light curve of individual kernels was extracted. }
   {The EUV flare kernels were small (
   $\lesssim 60~\text{pixels} \approx 1~\text{Mm}^2$) and a significant fraction were unresolved at a plate scale of 135~km/pix, in this flare. Furthermore, we derived surprisingly short EUV kernel heating times of less than a few seconds. The average profile exhibits a sharp rise of $1.7\pm0.3$ s from half-maximum, requiring an additional $2.3^{+0.7}_{-0.4}$ s to return to its reference value.}
{Our findings indicate that approximately half of the kernels were unresolved in this flare, despite the enhanced angular resolution offered by Solar Orbiter's proximity to the Sun at 0.38 AU here. Furthermore, we show that energy was only injected in a localised region ($\lesssim 1~\text{Mm}^2$) of flare ribbons for less than a few seconds. These results necessitate an in-depth investigation into the implications of such small-scale and transient injections on the energy flux deposited in solar flares, and the resulting response of the solar atmosphere.}


   \maketitle
%

\section{Introduction}

Flare ribbons form as a result of the deposition of energy in low layers of the solar atmosphere. By studying the dynamics of flare ribbons, one obtains an indirect measurement of reconnection in the corona. Flare ribbons can therefore be used as a probe of the fundamental energy release and particle acceleration processes ongoing in flares. Several works have studied the spatiotemporal evolution of flare ribbons in Ultraviolet/Extreme Ultraviolet (UV/EUV) \citep[e.g.][]{Fletcher+2004SoPh..222..279F, Fletcher_2009A&A...493..241F, Kazachenko_2017ApJ...845...49K} and hard X-ray \citep[e.g.][]{ 2003ApJ...595L.103K, KRUCKER20051707, Fivian+2009ApJ...698L...6F}. The Transition Region and Coronal Explorer (TRACE) provided a dataset of high-cadence (2~s) flare observations. TRACE had a $2''$ Point Spread Function (PSF) at 1 AU. Using 1600~\r{A} TRACE data, \cite{Fletcher+2004SoPh..222..279F} tracked the evolution of chromospheric ribbon kernels and found that the intensity of UV emission was correlated with the rate of coronal reconnection assuming a 2D reconnection geometry. In a later work, \cite{Fletcher_2009A&A...493..241F} identified conjugate footpoint sources by correlating the time profiles of UV kernels. They concluded that correlated pairs were, in fact, magnetically linked and that the time delays between conjugate pairs were consistent with heating by relativistic particle beams in the majority of cases. The Atmospheric Imaging Assembly (AIA) data have also been frequently used in UV/EUV flare ribbon studies \cite[e.g.][]{Naus_2022, Purkhart+2025A&A...698A.318P, Qiu+2025SoPh..300..137Q}. AIA has an angular resolution of $1.5''$ and a cadence of 12/24~s for EUV/UV channels, respectively \citep{Lemen_2012}. Although the aforementioned observations provide insights into flare ribbon dynamics, they are often saturated during flares and typically lack a combination of high-resolution observations at high-cadence.

In addition, X-ray flare observations are often limited by the instrument's imaging dynamic range and sometimes also by the time resolution. The Reuven Ramaty High-Energy Solar Spectroscopic Imager (RHESSI) effectively had a 4~s temporal sampling due to periodic rotation about its own axis on that timescale and an angular resolution $> 2.3 ''$ \citep{Lin+2002SoPh..210....3L}. Since 2020, the Spectrometer Telescope for Imaging X-rays (STIX) onboard Solar Orbiter has observed solar flares at radial distances as close as 0.3 AU \citep{Krucker_2020}. STIX observes flares with dynamic time-binning, which can reach a cadence of 0.3~s during large flares. The high-cadence data enables one to study rapid variations in X-ray emission during flares \citep[e.g.][]{Collier+2023, Collier+2024}. \cite{Ryan+2025A&A...703L..12R} even managed to image the non-thermal X-ray sources at 1~s cadence for an unusually bright estimated-X16.5 GOES-class flare. However, similarly to its predecessor, RHESSI, the instrument's capabilities in resolving small and faint structures are limited by a low dynamic range and angular resolution. Thus, it is informative to combine the diagnostic of high energy particles obtained from X-ray observations with high-resolution imaging data in order to gain insight into the spatiotemporal evolution of energetic particle energy deposition and heating in the solar atmosphere. 

In the past few years, the Interface Region Imaging Spectrograph (IRIS) has been operating in high-cadence mode during dedicated campaigns \citep{DePontieu+2014SoPh..289.2733D}. This has led to several important discoveries \citep[e.g.][]{French+2021ApJ...922..117F, Lorincik+2025ApJ...986...73L, Ashfield+2025NatAs.tmp..221A}, including the first clear evidence of super Alfvénic slipping reconnection using $2~\text{s}$ cadence 1330~\r{A} data from the Slit Jaw Imager \citep[SJI;][]{Lorincik+2025NatAs...9...45L}. Furthermore, \cite{Lorincik+2025ApJ...986...73L} demonstrated the utility of high-cadence IRIS spectroscopy as a  tool to identify the progression of heating at various heights in the transition region and chromosphere.

\newcounter{mycounter}
\setcounter{mycounter}{2} 

In addition to space-based EUV telescopes, ground-based telescopes have contributed greatly to the study of small-scale features in flare ribbons \citep[e.g.][]{Sharykin+2014ApJ...788L..18S, Thoen_Faber+2025A&A...693A...8T, Yadav+2025}. The 1.6~m Goode Solar Telescope (GST) has provided high-resolution of flare ribbons in the near infrared and optical at diffraction limits of $0.08''$ at 500~nm and $0.16''$ at $1~\mu{\text{m}}$ \citep{Polito+2023ApJ...944..104P}. Those data led to the discovery of distinct characteristics of the leading edge of flare ribbon kernels \citep[e.g][]{Xu_2016, Xu_2022} and to small-scale features in the magnetic geometry during flares \citep[e.g.][]{Sharykin+2017ApJ...840...84S, Xu+2018NatCo...9...46X}.

More recently \cite{Thoen_Faber+2025A&A...693A...8T} used high-resolution chromospheric data from the Swedish Solar Telescope (SST) to examine the fine-structures of flare ribbons. With the CHROMospheric Imaging Spectrometer (CHROMIS), they identified ``blobs'' in the $\text{H}_\beta$ channel with Full-Width at Half Maximum (FWHM) values between 140--200~km. Chromospheric blobs were also identified in the $\text{H}_\alpha$ and Ca \Roman{mycounter} 8542 \r{A} lines, indicating that the structures spanned multiple heights throughout the chromosphere. At the time the data were obtained, CHROMIS completed a cycle of wavelength sampling in 7~s. Remarkably, the pixel scale of CHROMIS is $0.0379''$. The authors concluded that the identified chromospheric blobs were fully resolved in the observation window. Their results were further corroborated by a second study that examined small-scale structures that extend vertically as plasma columns from the flare ribbon, termed “riblets”, for three flares using SST observations. They identified riblets with widths ranging from 110--310~km and vertical lengths between 620--1220~km \citep{Faberb+2026A&A...705A.174T}. The authors proposed a link between the observed features and substructures in the reconnected current sheet that led to fragmented energy release. 

Similarly, the 4~m Daniel K. Inouye Solar Telescope (DKIST) provides high-resolution flare observations. Recent work by \cite{Yadav+2025} identified chromospheric ribbon blobs in the red wing of the Ca \Roman{mycounter} line measured by the Visible Spectro-Polarimeter (ViSP) with sizes between 320--455~km. These were found to be significantly hotter than the surrounding ribbon structures. ViSP scanned the entire Field of View (FOV) in 3.11 minutes with a slit step size of $0.107''$. The Visible Broadband Imager (VBI) offers a higher spatial and temporal resolution than ViSP, at 9~s cadence and a spatial sampling of $0.01''$ at the $\text{H}_{\beta}$ line. However, the blobs were not identifiable in VBI images for that observation, possibly because VBI measures emission over a wider range of wavelengths than ViSP. In related work, \cite{Tamburri+2025} quantified the widths of post-reconnection flare loops using data from the DKIST and found that the mean loop width at the top of the arcade was 48.2~km. Together, these findings highlight the small spatial scales involved in flare processes.  

From a modelling perspective, various authors have mapped substructures in the flare current sheet to flare ribbon features. \cite{Wyper+2021ApJ...920..102W} used a static analytical model to trace plasmoids that form due to the tearing mode instability in the flare current sheet to ribbon structures. They demonstrated that the presence of plasmoids leads to spiral-shaped features along the ribbons. In addition, \cite{Dahlin+2025ApJ...993...31D} built on this work with a 3D magnetohydrodynamic (MHD) simulation of an erupting flux rope. By identifying regions with strong guide field in the current sheet, the authors isolated plasmoid structures and mapped these to synthetic ribbon observations. They further demonstrated the dynamics of spiral-like structures in flare ribbons that correspond to plasmoids in the corona. By providing a testable mapping between plasmoids in the current sheet and observable ribbon features these works enable us to infer the features of plasmoids in coronal current sheets using high-cadence, high-resolution observations of flare ribbons. In recent work, \cite{French+2025} studied the evolution of EUV ribbon kernels using flare campaign data taken by the High Resolution Imager (\HRIEUVNOSPACE) of Solar Orbiter's Extreme Ultraviolet Imager (EUI) and related their spatial scales to the growth and coalescence of plasmoids in the current sheet. The authors concluded that the characteristics were consistent with the predictions of the tearing mode. 

In this paper, we build on these publications by studying the spatial and temporal scales of EUV flare ribbon kernels at high-cadence and short-exposure using data from the Solar Orbiter major flare campaigns. In particular, the aim of this work is to quantify the temporal and spatial scales of energy deposition in a solar flare using these data obtained during the impulsive phase of a flare when particle acceleration is strongest. The unprecedented dataset offers non-saturated observations of EUV ribbons at a plate scale of 135 km/pix ($0.492''$/pix at 0.381 AU) and 2~s temporal cadence. Compared to the aforementioned works, the uniqueness of the \HRIEUV dataset lies in the combined temporal and spatial capabilities. The resolution of the EUV instrument was enhanced because Solar Orbiter was near perihelion during the observational window. In addition, \HRIEUV was operating in flare campaign mode, which involves a cycle of six short-exposure (0.04~s) frames followed by a normal exposure (2~s). This approach unveils the detailed structure of flare ribbons that are normally saturated. Other existing publications that have used the Solar Orbiter major flare campaign data include \cite{Young+2025ApJ...986...64Y, Tan+2025A&A...702A.189T, Tan+2025A&A...702A..88T, Yuhang+2025ApJ...985L..12G}. 

An overview of the observations and the data preprocessing steps is given in Sections \ref{sec:observations} and \ref{subsec:dataset_prep}, respectively. In Section \ref{subsec:watershed}, a classical computer vision method that was used to identify individual kernels along flare ribbons in EUV frames is detailed. From this, the size distributions and the average time profile of EUV kernels were derived and the results are presented in Section \ref{sec:results}. These findings have implications on flare modelling approaches that rely on kernel area measurements to estimate the energy flux of the injected particle beam. These models also rely on estimates of pulse duration to determine the time in which energy is injected. The implications are further explored in Section \ref{sec:discussion}.

\section{Observations}\label{sec:observations}

\begin{figure*}
    \centering
    \includegraphics[width=\textwidth]{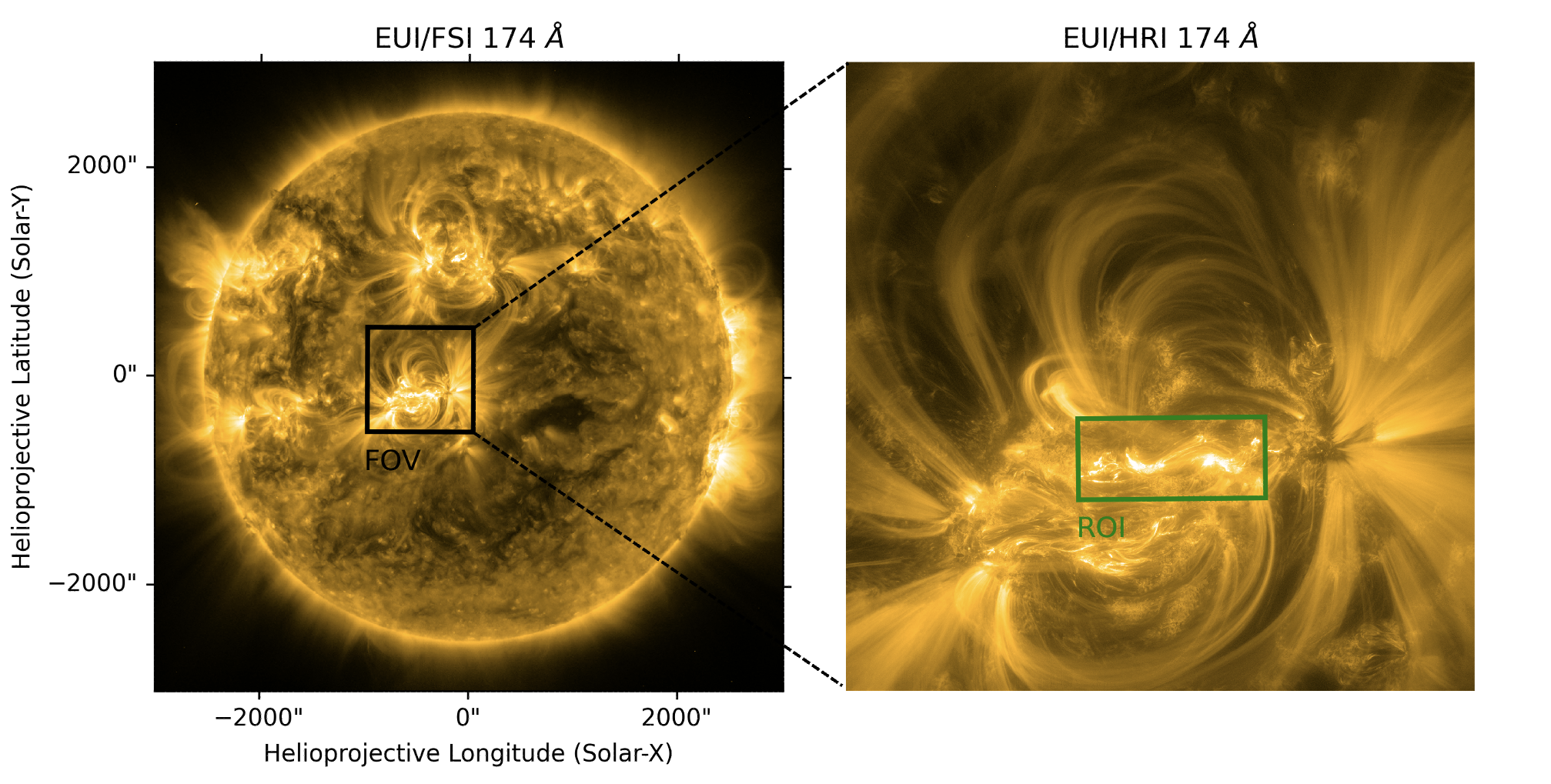}
    \caption{The flaring active region observed by FSI and \HRIEUV during the Solar Orbiter flare campaign on 2024-03-23 in log scale. The image on the right highlights the field of view of \HRIEUV and the green box denotes the region of interest within the FOV of HRI that is studied in this work. The \HRIEUV frame shows a normal exposure observation of the flare at the non-thermal peak time (23:41:14 UT). During the M2.5 GOES-class flare, the two bright flare ribbons were saturated in the normal exposure frames.}
    \label{fig:230324_ROI}
\end{figure*}

\begin{figure*}\label{fig:overview_ribbons}
    \centering
    \includegraphics[width=\textwidth]{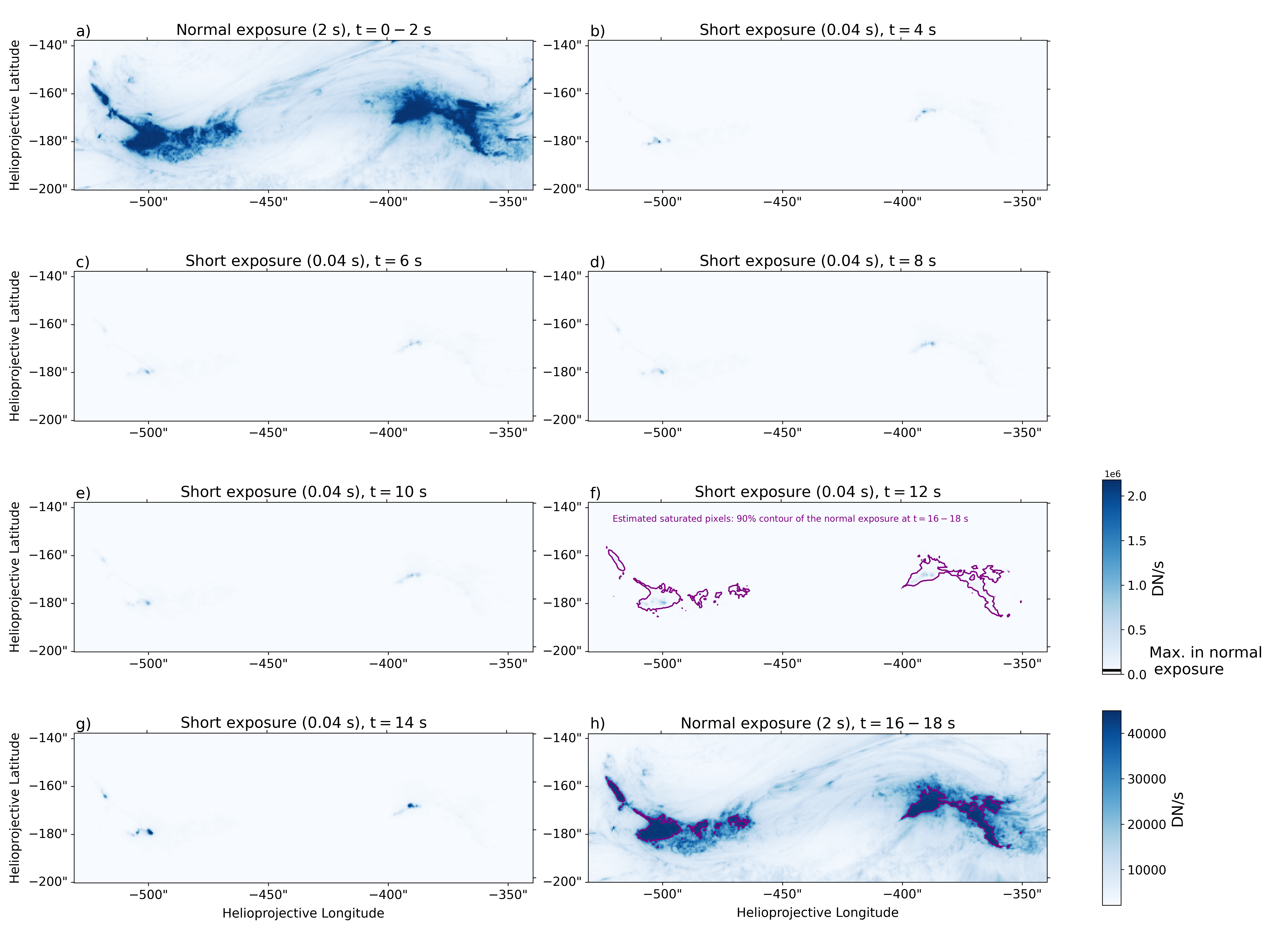}
    \caption{A normal exposure frame taken at 2024-03-23 23:40:10 UT ($
\text{t}=0$) followed by six short-exposures and finally a normal exposure when the cycle repeated. The first and last frames highlight the saturation levels reached in the active region during the main flare energy release. The short-exposures highlight the small-scale of EUV kernels within the otherwise fully saturated flare ribbons. All colourmaps are linearly scaled, with the short-exposure frames normalised to the maximum intensity at $\text{t} = 14~\text{s}$ and the normal exposures scaled to the saturation value. The associated movie is available online.}
\end{figure*}

The M2.5 GOES-class flare SOL2024-03-23T23:41 was observed by multiple Solar Orbiter remote-sensing instruments as part of the spacecraft’s major flare SOOP campaign \cite[see][]{2025SoPh..300..152R}. In particular, it was observed by the EUI instrument suite. The Full Sun Imager (FSI) observed the full solar disk in synoptic mode, and \HRIEUV observed the active region in a high-cadence flare-focused mode. During this time, Solar Orbiter was $3.33^{\circ}$ separated from the Sun-Earth line and at a distance of only 0.381 AU from the Sun. The target active region was located towards the South-East of the centre of the solar disk, at approximately (-480, -175)$''$, in helioprojective Cartesian coordinates from Solar Orbiter's viewpoint. Thus, absorption by structures along the line-of-sight was minimal, and those data are suitable for flare ribbon analysis. 

The observing mode for \HRIEUV involved a sequence of 6 short-exposure frames (0.04~s) followed by a normal exposure frame (2~s) taken at 2~s cadence. The readout time was approximately 1.35~s, which means that after a normal exposure was taken, there was an additional gap before the next cycle began \cite[see Figure 2 of][]{2025SoPh..300..152R}. For more details on the major flare campaigns and the EUI short-exposure mode, we refer the reader to \cite{2025SoPh..300..152R} and \cite{Collier+2024A&A...692A.176C_FSI}, for each topic respectively. The Field of View (FOV) of \HRIEUV during SOL2024-03-23T23:41 is marked by the black box on the FSI observation in Figure \ref{fig:230324_ROI}. The flaring region studied in the work is highlighted by the green box on the \HRIEUV normal exposure observation taken at the non-thermal peak time (23:41:14 UT --- all times reported in this paper are at Solar Orbiter). 

Figure \ref{fig:overview_ribbons} shows a normal exposure taken at 23:40:10 UT followed by six short-exposures and finally a normal exposure  when the cycle repeated. With the normal exposure time setting, the entirety of the flare ribbons were saturated and all details regarding the substructure of the ribbons were lost (panels a) and h)). The short-exposures (panels b)--g)) revealed the detailed distribution of emission in the ribbons. The kernels were remarkably small (a few pixels), indicating that a large fraction of energy was deposited in concentrated regions within the flare ribbons and active region. The exact kernel sizes were quantified and are later presented in Section \ref{sec:kernel_areas}. The extreme range in EUV intensity of the kernels is also highlighted. A peak intensity just over $2\times10^6~\text{DN}~\text{s}^{-1}$ was reached at $\text{t} = 14~\text{s}$, whereas the normal exposures saturated just above $4\times10^4~\text{DN}~\text{s}^{-1}$. This further reaffirms that flare energy release and deposition is highly inhomogeneous and that observations of the detailed sub-structures of flare ribbons promise to lead to significant advances in our understanding of the physics of flare processes.

\begin{figure}
    \centering
    \includegraphics[width=\columnwidth]{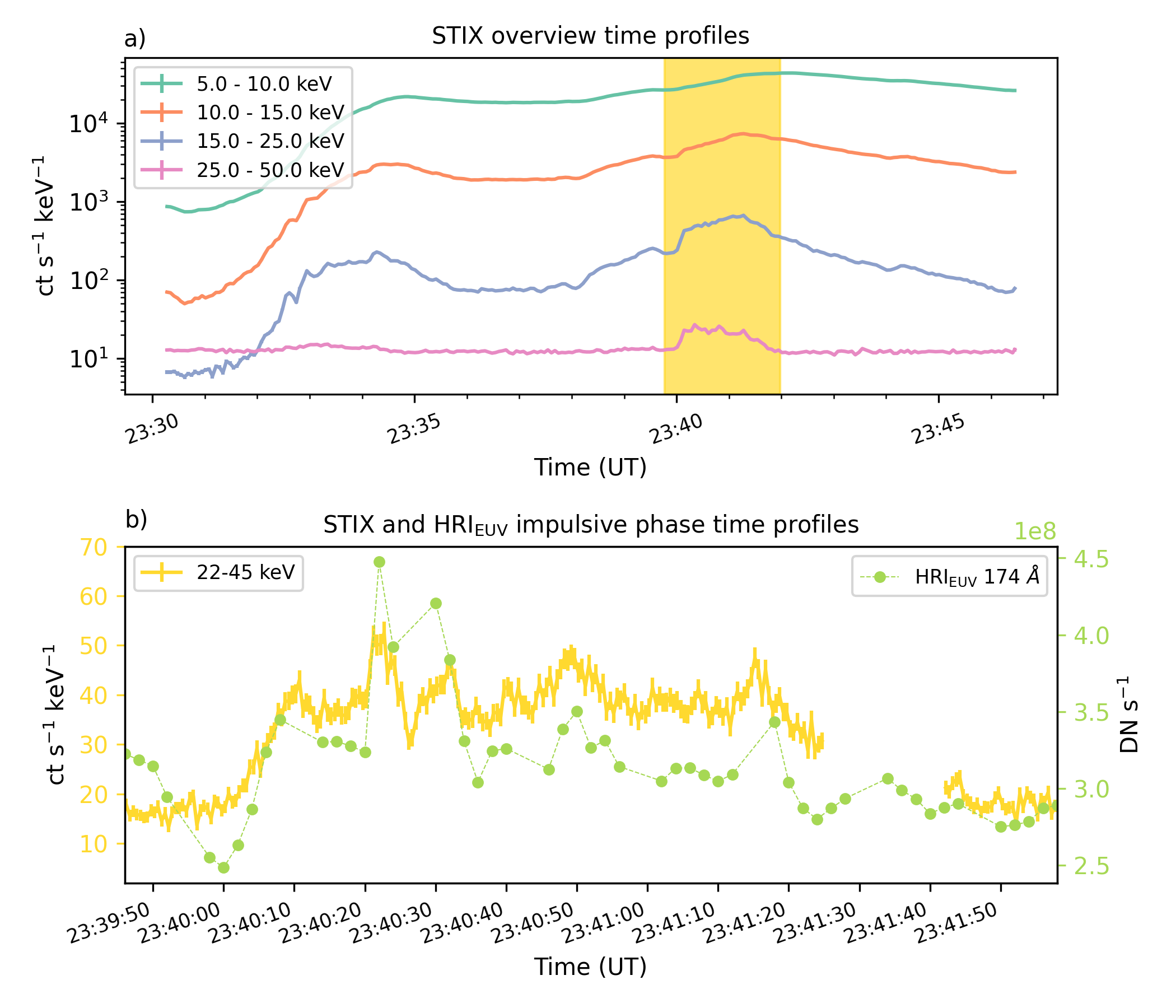}
    \caption{Overview time profiles of the M2.5 GOES-class flare SOL2024-03-23T23:41. Panel a) shows the STIX light curves for the entire flare and panel b) shows a closer look at the high energy X-ray profiles (22-45 keV) compared to the 174 \AA\ spatially integrated (over the ROI shown in Figure \ref{fig:overview_ribbons}), short-exposure time profile. The short-exposure EUV data was taken at $2~\text{s}$ cadence with $6~\text{s}$ gaps every 6 frames. The correspondence between hard X-ray and EUV emission is notable.}
    \label{fig:230324_overview_fig}
\end{figure}

Figure \ref{fig:230324_overview_fig} shows an overview of the EUV and hard X-ray time profiles of the flare event. Panel a) shows the STIX time series in the native quick-look energy binning, and panel b) highlights the main impulsive phase, which is coloured yellow in panel a). This represents the duration of interest studied in this chapter. The spatially integrated 174 \AA\ short-exposure time profile is shown in green (for the region denoted by the green box in Figure \ref{fig:230324_ROI}). Due to the uneven sampling cadence resulting from the interleaving data-taking approach of \HRIEUVNOSPACE, peaks in the EUV time profiles sometimes appear slightly offset with respect to the hard X-ray emission. STIX data were obtained at 0.5~s during the time range shown in panel b). The break in the STIX light curve between 23:41:24 UT and 23:41:42 UT is due to a data gap. We note that there is strong correspondence between the hard X-ray (22-45 keV) time profile and the short-exposure EUV time profile during the impulsive phase. This indicates that hard X-ray emission, that occurs as a result of electron energy deposition, radiated simultaneously with prompt emission in the 174~\r{A} passband.

\section{Methods} \label{sec:methods}
\subsection{Dataset preparation} \label{subsec:dataset_prep}
In this analysis, short-exposure Level-2 data provided by the EUI instrument team were used\footnote{Data release 6.0: \url{https://doi.org/10.24414/z818-4163}}. The data processing from Level-1 to Level-2 includes a correction to the absolute pointing of the spacecraft that was calculated on the basis of fitting the position of the solar limb in an observation by FSI close in time to the time range of interest. In addition, the measured data numbers were scaled to the high-gain values and normalised by the exposure time in the Level-2 dataset. In an additional step, a jitter correction was applied to the Level-2 data by aligning each short-exposure frame to a reference frame in the sequence (in this case the first frame) using \code{sunpy's} reproject functionality \citep{sunpy_community2020}. This is necessary due to known variations in the pointing of the spacecraft. An additional unknown jitter is sometimes present in the spacecraft pointing that is not captured by the as-flown attitude information. The removal of this jitter is usually achieved by co-alignment to stable features \citep[see e.g.][]{Zhong+10.1093/mnras/stac2545, Chitta+2022A&A...667A.166C}, which can be a challenge for short-exposure observations because of the lack of suitable bright, stable, and persistent features. However, for the study period, the issue of the residual jitter is $< 1~\text{pixel}$ (see the supplementary movies) and does not impact the analysis presented here. 

In this work, the focus is on bright emission from the flare ribbons, and therefore pixels with counts less than a certain intensity were masked and set to zero. Here, we chose a certain percentage of the maximum intensity of all images in the sequence. This removes contributions from non-ribbon sources and from pre-flare ribbon heating. Various thresholds were tested. Ultimately, thresholds that correspond to 5\%, 10\%, 20\%, 30\% and 50\% of the maximum value in the image sequence were applied. Depending on the purpose of the analysis, different threshold choices are appropriate. For estimating kernel sizes, a lower threshold is suitable, since a high threshold artificially reduces the size of the kernels. Whereas for extracting light curves of a single event, a higher threshold minimises the contribution from multiple kernels. Furthermore, in order to identify newly brightened kernels or those that have become brighter since the previous frame, a ``running difference'' was applied to the data. To do this, the previous frame was subtracted from each image and all negative values were set to zero. This approach identifies kernels that have become brighter since the previous frame and filters out those that have not.

\subsection{Kernel identification} \label{subsec:watershed}

Ribbon kernels, defined as bright spots located along flare ribbons, were best identified using a classical image segmentation method called \texttt{watershed} \citep[e.g.][]{Watershed1, Watershed2}. For comparison, the random walker  segmentation method\footnote{\href{https://scikit-image.org/docs/stable/api/skimage.segmentation.html\#skimage.segmentation.random_walker}{Scikit-image \texttt{random walker} documentation}} was also tested, but due to steep gradients in the data, the method did not perform as well as watershed. We therefore proceeded with the watershed approach in our analysis. Watershed is a well established segmentation method that excels at finding the boundaries of objects that overlap or touch. It does this by simulating topological flooding; from user given markers, the algorithm grows regions by following the path of steepest gradient until an edge or another object is encountered. In a first step, we applied the Euclidean Distance Transform (EDT) to each map that finds the distance of each foreground pixel to the background. This provides an approximate gradient field of the image that is used to cleanly segment kernels. We then find the local maxima in each EDT image. In doing this, a minimum marker separation of 4 pixels was used. This choice was made based on our current understanding of the \HRIEUV PSF. Therefore, in our analysis, a kernel is specifically defined as having a local maximum in intensity that is at least 4 pixels away from nearby bright spots. The algorithm then ``floods'' the region of interest at the location of each marker) until it reaches the background or other previously identified kernels. In order to apply this method, functionality from the Python software package scikit-image was used\footnote{\href{https://scikit-image.org/docs/0.25.x/api/skimage.segmentation.html\#skimage.segmentation.watershed}{Scikit-image \texttt{watershed} documentation}}. 

\begin{figure}
    \centering
    \includegraphics[width=\columnwidth]{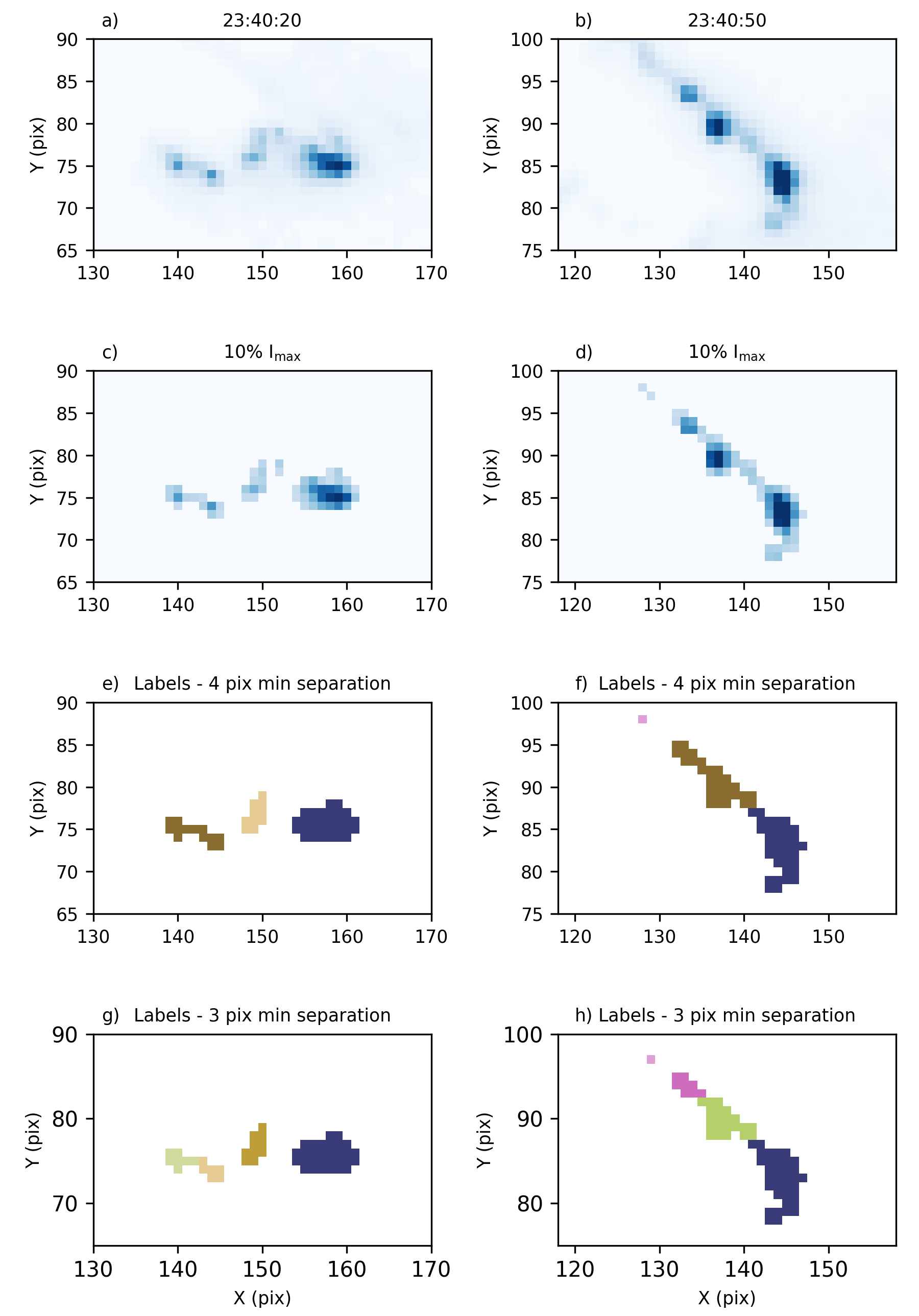}
    \caption{Examples of kernels identified using the watershed method in two frames at 23:40:20 UT and 23:40:50 UT. The first row shows the raw data measured by \HRIEUVNOSPACE. The second row shows the data after the $10\%~\text{I}_{\text{max}}$ threshold has been applied. Rows three and four show the labels that were output by the watershed segmentation method when a minimum marker separation of 4 and 3 pixels was used, respectively.}
    \label{fig:example_small_kernels}
\end{figure}

Figure \ref{fig:example_small_kernels} shows an example of the watershed algorithm applied to EUV kernels. Sections of the EUV ribbons at two times, 23:40:20 UT and 23:40:50 UT, are shown. The second row shows the ribbons after the 10\% threshold has been applied. The third row shows the segmentation labels that were output by the watershed algorithm when a minimum marker separation of 4 pixels was used. For comparison, we also show the segmented ribbons when a marker separation of 3 pixels was used. These are examples of EUV ribbon kernels that were not resolved by the instrumental PSF. The segmentation worked best on these examples when a 3 pixel separation was used, however, in others cases it caused over-segmentation of ribbons. Overall, the results presented in Section \ref{sec:results} were largely consistent when the two approaches were compared.

The efficacy of this method depends primarily on the reliability of the markers provided by the user. The approach of obtaining markers by finding local maxima in the EDT image works best for circularly shaped structures rather than elongated structures. In our dataset, the brightest kernels tend to be approximately circularly shaped, however, oftentimes the regions in between the brightest part of the ribbons are rather elongated. This may lead to splitting of elongated structures, depending on the minimum marker separation allowed. Another approach is to provide markers by finding the local maxima in the original data itself, rather than the EDT. For comparison, both approaches were tested. We found that the results presented in Section \ref{sec:results} did not vary drastically, despite slightly different kernel identifications in some frames. Although the regions between the brightest kernels may visually appear too faint to be classified as kernels, we note that the data was taken in only a 0.04~s exposure. Although their relative brightness is low compared to the brightest kernels, those pixels were, in fact, rather bright sub-structures of EUV flare ribbons and are therefore classified as ribbon kernels.

Previous works that have performed segmentation of magnetic structures in the solar atmosphere, have tracked the structures across multiple frames and have used overlap criteria to link kernels in different frames. The linked kernels were considered to be the same structure. Four established methods were compared by \cite{DeForest_2007ApJ...666..576D}. The authors showed that the results differ, particularly for low flux data or data with a low signal-to-noise ratio. In this work, no tracking was performed. The kernels were treated as independent structures from frame to frame. We took this approach for various reasons. Firstly, we know that the energy release in flares is highly dynamic and we hypothesise and later show that heating only occurred in a localised region of space for a very short time. Secondly, in order to  track kernels, assumptions must be made regarding the degree of overlap that warrants kernel merging. This is not trivial, particularly, given that we later show, that a significant fraction of kernels were unresolved at this angular resolution. As a result, it is challenging to reliably determine whether two overlapping kernels are truly from the same coronal reconnection event or are simply overlapping structures due to the resolution offered by the instrument at this radial distance.

\section{Results}\label{sec:results}

\subsection{Kernel areas} \label{sec:kernel_areas}
\begin{figure*}
    \centering
    \includegraphics[width=\linewidth]{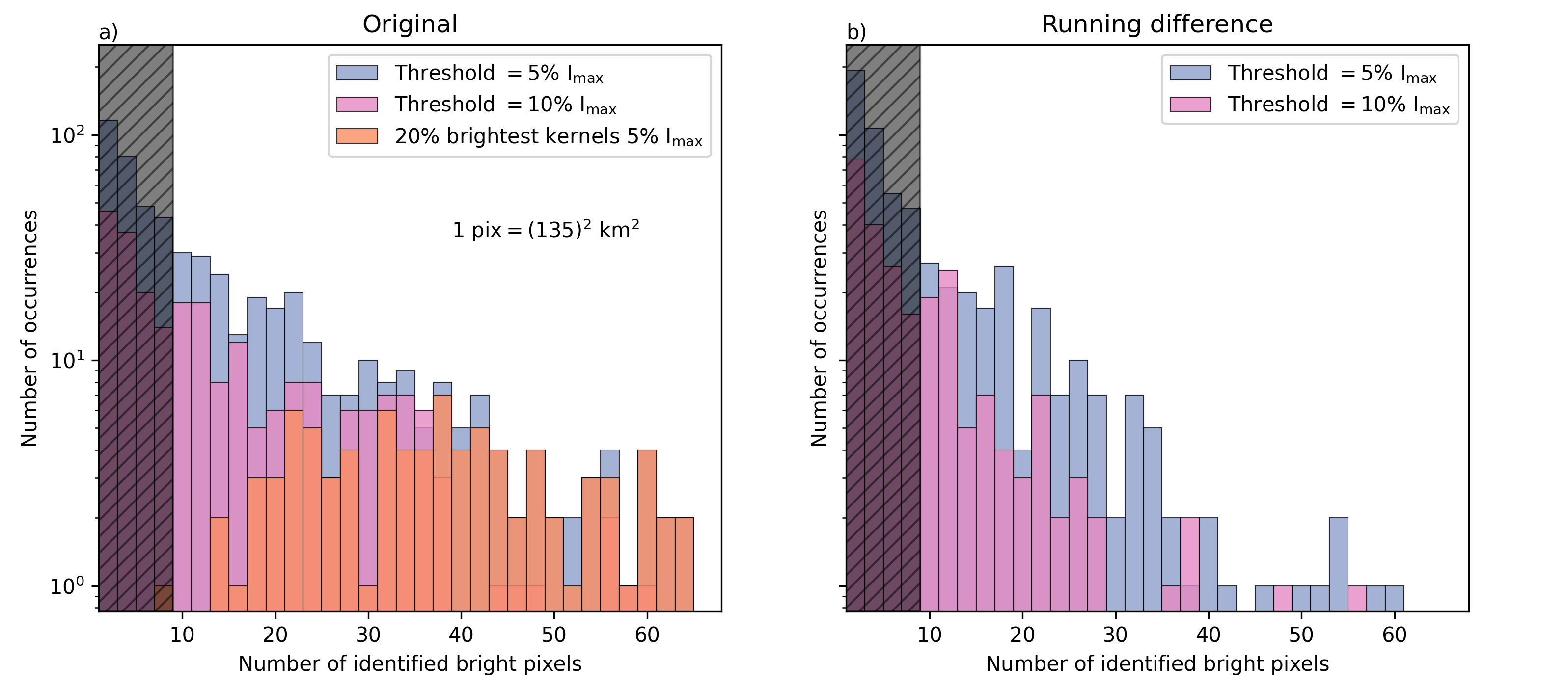}
    \caption{Kernel size histograms showing the number of bright pixels identified in a kernel using the watershed method with two different thresholds applied to both the original and the ``running difference'' data. These distributions demonstrate that a significant fraction of kernels were unresolved and that the brightest kernels had sizes that were on the order of the instrument's PSF. The characterisation of faint kernels depends strongly on the threshold applied and their sizes are thus somewhat ambiguous. The vertical shaded area masks kernels whose size estimates are unreliable due to artificial shrinkage caused by data thresholding.}
    \label{fig:kernel_areas_histogram}
\end{figure*}

Using the EUV kernel pixels identified by the watershed algorithm, their sizes were quantified. These constraints are vital for calculating flare energy flux, a quantity that is sensitive to the estimated cross-sectional area of the injection site. Figure \ref{fig:kernel_areas_histogram} panels a) and b) show histograms of the kernel sizes obtained when 5\% and 10\% thresholds were applied to the original and ``running difference'' data, respectively. The distributions peak at the smallest kernel sizes and decrease sharply with increasing size. It is also notable that there is an excess of kernels that contain between 25--60 pixels in the original data distributions compared to those based on the ``running difference" data. This is likely because kernels appear to be larger in the original data due to pre-existing emission in the region, whereas with the ``running difference" method pre-existing emission is removed. 

\HRIEUV was designed for pixel-limited resolution \citep{EUI_instrument_paper} and early observations \cite[Fig. 24,][]{Berghmans+2023A&A...675A.110B} indeed indicate that the sharpest observable features are close to the Nyquist Sampling limit of 2-pixels. In Appendix \ref{appendix:2pix_psf}, Figure \ref{fig:point_source_example}, an example is given showing a point source smeared out by a PSF with a 2-pixel FWHM. The point source is centred in the middle of a pixel and at the intersection between four pixels in the top and bottom row, respectively.  If the thresholds are not chosen well, the method applied in this work can lead to kernel shrinkage. This effect is demonstrated in the final four columns of Figure \ref{fig:point_source_example} where four cut-off thresholds were applied to a single kernel. It is clear that a point source with this PSF may appear to have anywhere between $\sim$ 1--20 pixels depending on the position of the centre of the source on the pixel and the cut-off threshold applied. Since the threshold is a fixed value, fainter kernels are more susceptible to artificial source shrinkage.  

Figure \ref{fig:kernel_areas_histogram} indicates that a significant fraction of EUV kernels were unresolved when observed with the instrumental resolution offered by \HRIEUV at this radial distance. To quantify this, we excluded kernels that contain less than 10 pixels from our analysis because we know that those are affected by artificial shrinkage. This amounted to 50\% of the total kernels when the 10\% threshold was applied to the original data. We estimated that structures with 10--20 pixels correspond to roughly 1 PSF (see Appendix \ref{appendix:2pix_psf}). Of the remaining 50\%, 50\% contained 10--20 pixels when segmented on the original data using a threshold of 10\%. The fraction decreased to 43\% of kernels when a 5\% threshold was applied. Using the ``running difference" data, the percentages increased to 73\% and 60\% when the 10\% and 5\% thresholds were applied, respectively. This highlights the fact that energy deposition occurred on scales unresolved by the instrument in a significant fraction of events. Figure \ref{fig:kernel_areas_histogram} a) also shows the kernel size distribution of the 20\% brightest kernels (based on their maximum intensity). The mean of the distribution is greater than that obtained when the entire range of kernel intensities was included and there is a turnover in the distribution at small kernel sizes. This difference is partly attributable to the effect of kernel shrinkage due to the application of thresholds to the data. In Appendix \ref{appendix:kernel_size_50pc} we also show the size distributions for a subset of kernels whose watershed labels extend beyond pixels with intensities equal to 50\% of the peak kernel intensity. The labels were clipped to only include pixels with intensities $\geq 50\% ~\text{I}_{\text{kernel,peak}}$. This removes any differences introduced by using a fixed threshold. See Figure \ref{fig:kernel_sizes_above_50p} for a comparison with Figure \ref{fig:kernel_areas_histogram}. Since the kernels identified in this dataset, are close to and sometimes at the resolution limit of the instrument, the source size estimates are even sensitive to where the source falls on the pixel grid, as demonstrated in Figure \ref{fig:point_source_example}. Kernels are not smoothly sampled and have steep gradients. As a result, it is challenging to determine the true source sizes and we therefore caution the reader not to over-interpret the shape of the distributions presented in Figure \ref{fig:kernel_areas_histogram} and Figure \ref{fig:kernel_sizes_above_50p}. The figures primarily serve to highlight the main range of kernel sizes detected by \HRIEUV in this dataset. We also note that a very small fraction ($\sim1\%$) of kernels detected contained more than $60$~pixels (not shown in Figure \ref{fig:kernel_areas_histogram}).

It is important to note that the kernel sizes reported here are simply the total number of pixels, despite the fact that the ribbons were sometimes identified as long, thin structures (see the supplementary movies) and, as such, were narrower in one dimension than in the other. As a result, stating whether a kernel was resolved or not based on the smearing of a point source due to the instrumental PSF is a simplified approach that does not always entirely represent the data. It could be the case that a kernel may be resolved in one direction but not in the other. Nonetheless, using this method in combination with a careful inspection of the data, we concluded that a significant fraction of identified EUV kernels were spatially unresolved in this flare. 

Importantly, these results place upper bounds on the cross-sectional areas in which the bulk of energy is deposited in the solar atmosphere during solar flares. These findings will increase the energy flux in typical electron beams and, therefore, has implications for flare modelling that will be discussed further in Section \ref{sec:flare_modelling_implications}. As of yet, it is not clear by how much the beam flux would increase on average. This finding warrants further investigation as well as the development of even higher resolution flare optimised EUV/UV telescopes. Additionally, continued coordinated flare observations with higher resolution ground-based telescopes are crucial to further constrain the spatial scales involved in flare energy release.

\begin{figure*}
    \centering
    \includegraphics[width=1\linewidth]{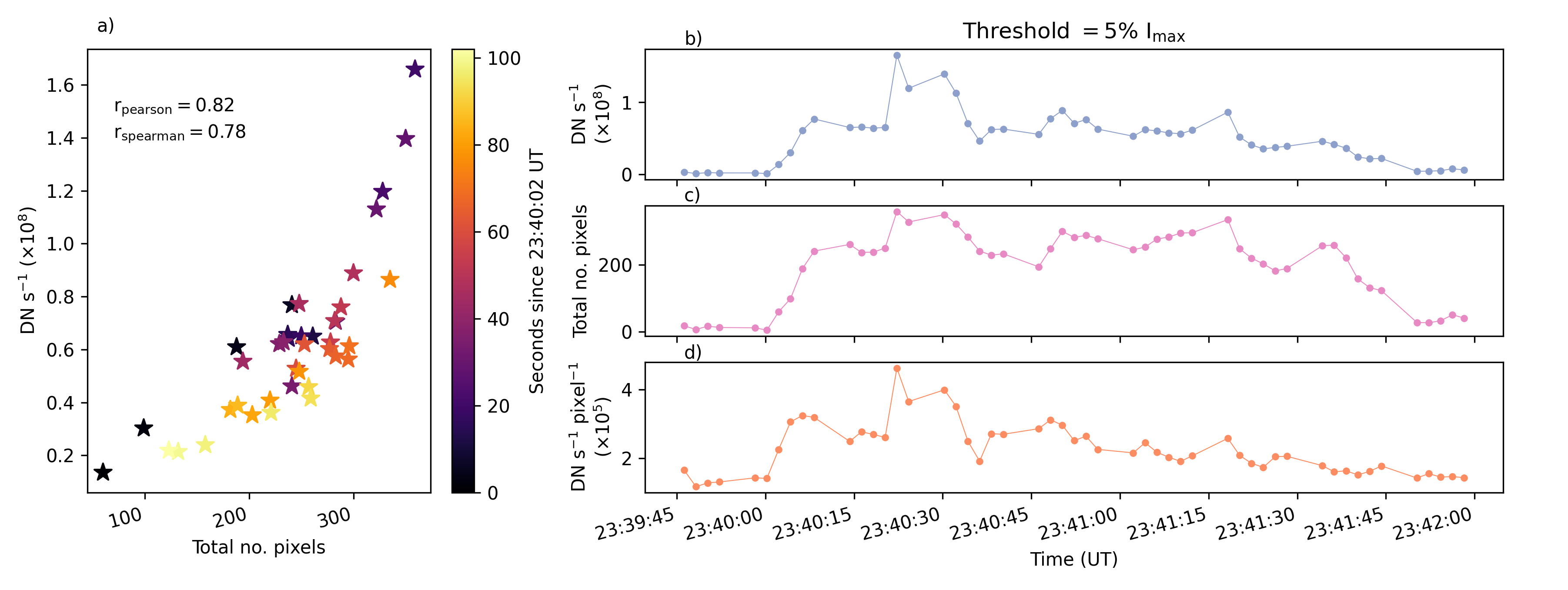}
    \caption{Instantaneous ribbon area versus the total EUV intensity in each frame. Panel a) shows that there is a positive relationship between the total instantaneous intensity of the EUV ribbons and the total instantaneous ribbon area. Panels b), c) and d) demonstrate that when normalised by the ribbon area, the flux has the same temporal variability as the original ribbon light curve.}
    \label{fig:brightness_vs_area}
\end{figure*}

Figure \ref{fig:brightness_vs_area} a) shows the total short-exposure EUV intensity (in $\text{DN}~\text{s}^{-1}$) versus the total number of bright pixels in each frame, identified by the watershed algorithm when the 5\% threshold was applied. There is a strong positive correlation ($\text{r}_{\text{pearson}} = 0.82, \text{r}_{\text{spearman}} = 0.78$, $\text{p} < 0.001$ for both) between the total intensity of the EUV ribbons (in $\text{DN}~\text{s}^{-1}$) and the total number of kernel pixels in a frame. In other words, the total instantaneous EUV ribbon intensity (shown in panel b)) correlates with the instantaneous ribbon area (shown in panel c)). This result is surprising because, if the relationship were perfectly linear, it would indicate that the EUV flux was evenly distributed across kernels in time. However, panel d) shows the total DN/s normalised by the instantaneous ribbon area. Interestingly, the time evolution corresponds precisely to the time evolution of the spatially integrated EUV ribbon intensity (panel b)). In other words, the temporal evolution is still present in the EUV flux. As a result, we conclude that the total ribbon intensity correlates with the ribbon area to a certain degree; however, there was also variation in the energy flux within certain regions, that led to the temporal evolution shown in panel d). 

It is evident from Figure \ref{fig:230324_overview_fig} that the spatially integrated EUV ribbon emission strongly correlates with the temporal evolution of the hard X-ray (22-45 keV) emission during the impulsive phase of the flare. This complements previous results that correlated $\text{H}_{\alpha}$ wing emission from flare ribbons with hard X-rays and from these measurements compared the reconnection rate with the electron energy release rate \citep[e.g.][]{Lee+2006ApJ...647..638L, Cannon+2023ApJ...950..144C}. By combining the results shown in Figure \ref{fig:230324_overview_fig} and Figure \ref{fig:brightness_vs_area},  we note that there is also a positive correlation between the instantaneous ribbon area at high-cadence and the hard X-ray emission  in this flare. It is also interesting to this relationship to previous work by \cite{Kazachenko_2017ApJ...845...49K}, who found a strong positive correlation ($\text{r}_{\text{spearman}} = 0.68 \pm 0.01$) between peak soft X-ray flux in flares and the cumulative ribbon area for the entire flare, for a dataset of more than 3000 C-X class flares. The result presented here differs from the relationship highlighted by \cite{Kazachenko_2017ApJ...845...49K} since here the instantaneous ribbon area is compared to the hard X-ray emission. Nonetheless, these relations could be causally linked due to the well-established relationship between the integral of the impulsive energy input to soft X-ray emission in flares, known as the Neupert effect \citep{Neupert_1968ApJ...153L..59N}. 

It is also worth noting how this builds on the current understanding of temporal variability in flare emission on short timescales, commonly referred to as short-period quasi-periodic pulsations \citep[QPPs;][]{nakariakov2009, vandoorsselaere_2016, mclaughlin2018,  Kupriyanova2020quasi-periodic, zimovets2021}. This finding demonstrates that each hard X-ray pulse corresponds to a brightening of a new part of the flare ribbons, in this event. However, the flux of that brightening appears to vary depending on the energetics of the reconnection process (the injected energy flux), which is characterised by hard X-ray temporal variability, both in total intensity and spectral index. 

\subsection{An average EUV kernel time profile} \label{sec:average_time_profile}

\begin{figure}
    \centering
    \includegraphics[width=\columnwidth]{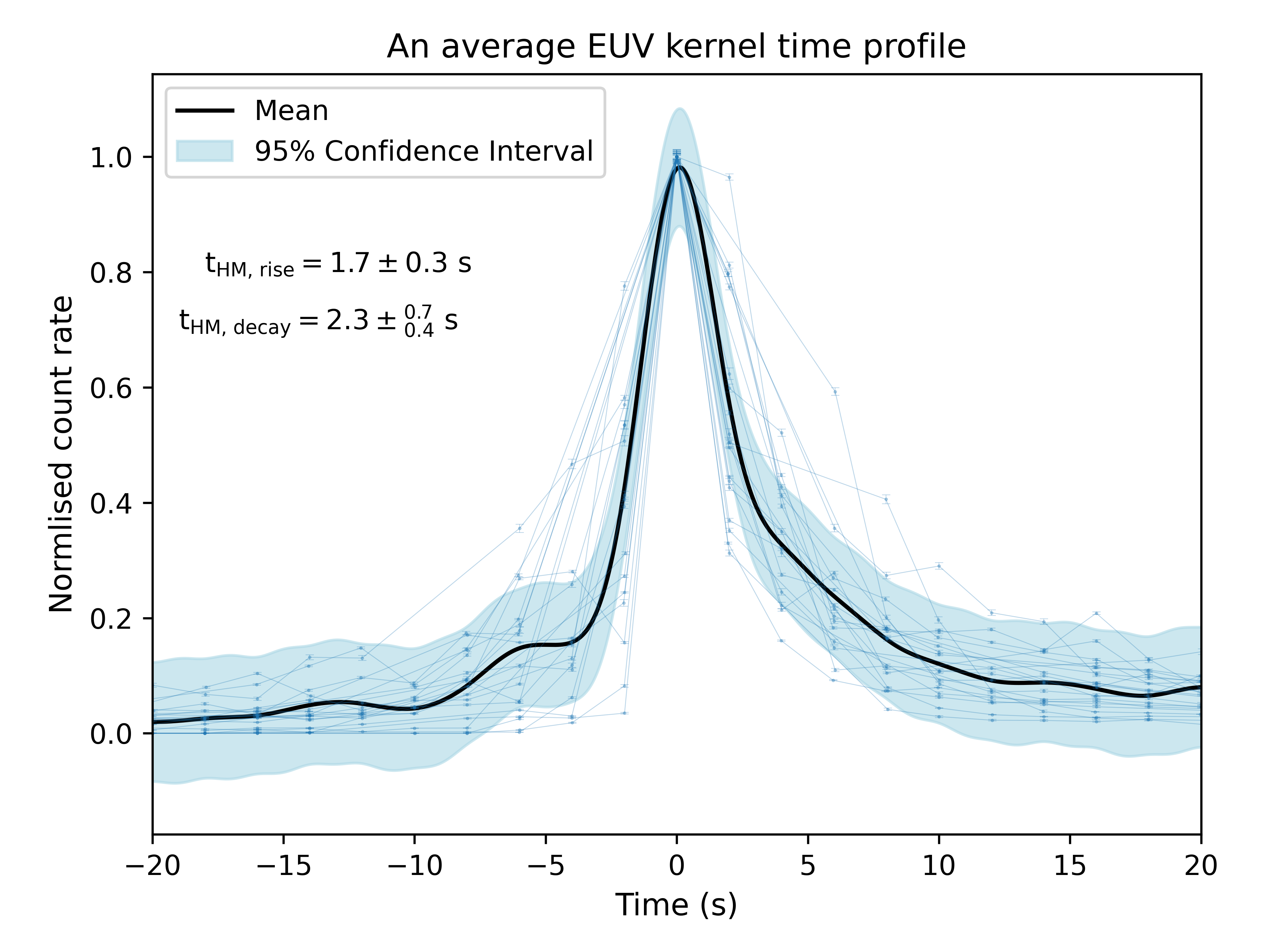}
    \caption{The normalised time profiles of 25 individual kernels fitted using Gaussian Process regression. To identify the kernels a threshold of 50\% $\text{I}_{\text{max}}$ was applied in order to exclude the fainter EUV kernel pixels. The lightcurves of individual EUV kernels are plotted below. The fit gives short rise and decay times of $\text{t}_{\text{half-max, rise}} = 1.7\pm{0.3}~\text{s}$ and $\text{t}_{\text{half-max, decay}} = 2.3^{+0.7}_{-0.4}~\text{s}$ respectively.}
    \label{fig:50pc_threshold_lc_gp}
\end{figure}

In the following subsection, an average EUV kernel time profile is presented. The aim of this task was to determine the dwell time in a spatially localised region of the ribbons. To do this, we extracted the light curves of each individual kernel identified using the watershed method, with the 10\%, 20\%, 30\%, and 50\%  cut-off thresholds applied. Several of the kernels that had been identified using the watershed method have common pixels with kernels identified in other frames. Those pixels were flagged and only included once when computing the average EUV time profile. If conjugate kernels located on opposite-polarity ribbons were associated with the same coronal reconnection event, their time profiles would be almost identical. This was the case in a few instances. However, those time profiles were not removed from the dataset. Figure \ref{fig:50pc_threshold_lc_gp} shows the time evolution of EUV kernels that were identified with the 50\% threshold applied. They are plotted as points with capped error bars that represent the measurement uncertainty connected by light blue lines. The time profiles have been normalised to their peak value and aligned so that they reach their maximum at $\text{t}=0$. Several light curves had multiple peaks, and those with secondary peaks above 0.3 were removed from the analysis. This resulted in 25 light curves. 

The data were fitted using Gaussian Process (GP) regression \citep{Rasmussen2004}. This approach was taken because it allows one to fit a non-parametric curve to data while taking into account the known uncertainty on the data and also estimating additional epistemic uncertainty. The known measurement uncertainty on \HRIEUV data is a combination of photon shot noise and sensor read-out noise. These noise contributions are described in the EUI user manual\footnote{\href{https://www.sidc.be/EUI/data/latest_release_notes.html}{EUI user manual}}. It is evident from the large spread in the data points at each time step in Figure \ref{fig:50pc_threshold_lc_gp} that epistemic uncertainty is the dominant source of uncertainty in our data. To fit the model, the data were scaled to have zero mean and unit variance using the scikit-learn StandardScaler functionality\footnote{\href{https://scikit-learn.org/stable/modules/generated/sklearn.preprocessing.StandardScaler.html}{Scikit-learn StandardScalar documentation}}. We then represent the data with a scaled Matern covariance matrix and a white noise contribution, which accounts for unknown noise sources. A Matern component was used instead of the commonly used radial basis function because although the desired fit should be smooth, there are sharp transitions around the peak that are better represented by a Matern function. Using a training dataset, the following optimal hyperparameters were found that give a log marginal likelihood, $\text{LML} = -360$: Matern kernel amplitude, $\text{A} = 0.706^2$, length scale, $\text{l}=0.06$, smoothing parameter, $\nu = 5/2$, and white noise level, $\sigma_{\text{wn}}^2 = 0.0918$. The known measurement uncertainties were taken into account by the $\alpha$ term that was added to the diagonal of the covariance matrix of the prior. They were small compared to the variance estimated by fitting the white noise contribution. In addition, the optimised Matern kernel length scale was close to the minimum spacing between data points in the rescaled units. This indicates that two data points become uncorrelated very quickly with increasing distance between them and the GP finds localised variations in the data. These variations may not be physically real and could be due to the large spread in measured y-values that is not fully represented by the white noise component. The source of this variation could be due to a temporal undersampling of the profile or, more speculatively, different heating sources producing a wide range of heating profiles \citep[e.g.][]{Kerr+2026}.  When assessed on a test set, the predicted values give a mean squared error, $\text{MSE} = 0.08$. The resulting fit is shown in Figure \ref{fig:50pc_threshold_lc_gp}.

The time for the average EUV kernel profile to rise from half-maximum to peak and decay again was calculated from the fit. The values obtained are $\text{t}_{\text{half-max, rise}} = 1.7\pm0.3~\text{s}$ and $\text{t}_{\text{half-max, decay}} = 2.3^{+0.7}_{-0.4}~\text{s}$. These estimates indicate that heating and cooling occurs on incredibly short timescales in localised bright kernels in flares. For comparison, we note that flare simulations that inject particle beams as the energy input into a flare loop typically model injections that last 10--20~s 
\citep[e.g.][]{Kerr+2022FrASS...960856K}. 

To support Figure \ref{fig:50pc_threshold_lc_gp}, Appendix Figure \ref{fig:average_lc_w_bkg} shows the time evolution of individual kernels in grey when the 10\%, 20\%, 30\% and 50\%  thresholds were applied. There are 179, 103, 62 and 37 time profiles for the four thresholds in increasing order. The light curves were produced using the same method as previously described, except that no time profiles were removed on the basis of the presence of multiple peaks in the time series. This may be a better approach since \cite{Dahlin+2025ApJ...993...31D} demonstrated that the presence of multiple peaks in a kernel time series could be due to the evolution of spiral-like features in flare ribbons related to the mapping of plasmoid structures from the current sheet to the flare ribbons and not necessarily due to multiple injections. Therefore, we extracted the average profile for both cases, i.e. both excluding (Figure \ref{fig:50pc_threshold_lc_gp}) and including (Appendix Figure \ref{fig:average_lc_w_bkg}) kernel time series with multiple peaks. For Appendix Figure \ref{fig:average_lc_w_bkg} an average profile was derived simply by taking the mean of each of the interpolated individual kernel profiles. In each panel, the average light curve is shown on top of the grey lines. The light curves are much noisier and have many more secondary peaks when lower thresholds were applied. This is possibly due to the fact that individual events are not resolved (see Figure \ref{fig:kernel_areas_histogram}), and therefore a superposition of events exists within a single kernel identification. In addition, in all cases, the profiles are evidently undersampled. Therefore, they provide an upper bound on the half-maximum rise time that is between 1.4--4~s. 

Estimating the average injection time is not trivial for various reasons. It depends on the choice of threshold used to identify kernel pixels. This effect is evident from Appendix Figure \ref{fig:average_lc_w_bkg}. In addition, there appears to be both a gradual and impulsive component to the heating profile. Therefore, choosing an appropriate reference background measure to use may vary depending on the application. Here, we used half-maximum as a reference point. Furthermore, the profile is so sharp that it is likely that the peak is frequently missed at a temporal cadence of $2~\text{s}$. Finally, the fact that we are not resolving all kernels spatially also suggests that we are not able to resolve their temporal evolution. These effects cumulatively lead to an overestimation of the FWHM of the average EUV kernel time profile. For all of these combined reasons, we conclude that the inferred half-maximum injection times reported here represent an upper limit on the true injection time in a localised region.

\section{Discussion} \label{sec:discussion}

\subsection{The response of HRI 174~\r{A} to impulsive heating}
The \HRIEUV instrument uses multi-layer mirror coatings and aluminium filters to isolate photons with wavelengths in a narrow range around 174~\r{A}. Nonetheless, the instrument has a broad temperature response, with contributions from $0.1-30~\text{MK}$ plasma, with the most prominent peak around 1~MK. This effectively means that \HRIEUV is most sensitive to plasma from the corona or heated plasma from the lower layers \cite[see Figures 2 and 3 of][]{HRI_response_2025A&A...699A...7S}. The flare ribbon kernels identified in this work were heated to this temperature range on extremely short timescales (several seconds) and correspondingly cooled on slightly longer but comparable timescales. This is plausible given impulsive injection of an energetic electron beams \citep[e.g.][]{Collier+2024A&A...692A.176C_FSI}. However, when hard X-ray imaging was performed for this event, a significant fraction of the hard X-ray emission appeared to originate from the region between the EUV ribbon. In addition, the electrons that emitted X-ray bremsstrahlung had very soft spectra (not shown). These analyses indicate that a large fraction of energetic electrons were trapped in the corona, possibly due to a high coronal target density \cite[like in e.g.][]{Veronig+2004ApJ...603L.117V, Veronig+2005AdSpR..35.1683V}. In addition, the event occurred during the decay of a long duration Solar Energetic Particle (SEP) event that struck the spacecraft following an X-class flare from earlier that day. SEP events contaminate the hard X-ray signal measured by STIX and thus complicate any hard X-ray imaging and spectral analysis \citep[e.g.][]{Collier+2024ITNS...71.1606C}. As a result, the analysis of the hard X-ray component of this flare is quite complex, and it is not clear whether electron beams were responsible for sufficient energy transport to the low solar atmosphere to account for the impulsive heating of the low corona measured by \HRIEUVNOSPACE. The complications in hard X-ray analysis are merely highlighted here; full analysis of the X-ray emission component is beyond the scope of this paper. Nevertheless, the clear temporal correlation between high energy hard X-ray time series (22-45~keV) and the spatially integrated EUV flux (shown in panel b) of Figure \ref{fig:230324_overview_fig}), strongly suggests that a dominant fraction of the hard X-ray flux was also coming from the ribbon kernels. 

\subsection{Implications on flare modelling}\label{sec:flare_modelling_implications}
Field-aligned 1D flare loop hydrodynamic models are often used to study the atmospheric response to energy injection from flares. Since cross-field diffusion of plasma is inhibited in the solar corona, the single strand approximation is justified. 1D models are less computationally intensive than those in 3D and as a result they serve as a useful tool to probe the physics ongoing at various altitudes of the solar atmosphere when faced with a large injection of energy. Some examples of 1D loop models include the radiation hydrodynamic codes RADYN \citep{carlson1992ApJ...397L..59C, carlson1995ApJ...440L..29C, carlson1997ApJ...481..500C, carlson2002ApJ...572..626C, abbett1999ApJ...521..906A, 2005ApJ...630..573A, 2015ApJ...809..104A} and HYDRAD \citep{Bradshaw+2003A&A...401..699B, Bradshaw+2003A&A...407.1127B}. For a review of 1D flare loop models, see \cite{Kerr+2022FrASS...960856K, Kerr+2023FrASS...960862K}. These models simulate the atmospheric response to various sources of energy input, including electron beams. The energy flux of an electron beam injected into a single loop is used as input to the simulation. Hard X-ray observations are used to derive the injected electron beam parameters.  From these the total power of the non-thermal electron beam (often given in units of $\text{erg}~\text{s}^{-1}$) is computed \citep[e.g.][]{Milligan+2014ApJ...793...70M}. This is further normalised by the cross-sectional footpoint area in which the energy is deposited to get the energy flux of the electron beam that is passed as input to these models. The cross-sectional area is usually calculated from hard X-ray contours or UV/EUV flare ribbon observations. However, the low spatial resolution of most telescopes leads to an overestimation of the cross-sectional area in which energy is deposited and, therefore, to an underestimation of beam energy flux. This is particularly problematic in hard X-ray wavelengths, and hence EUV/UV images are often used. However, in this work, we have shown that even with state-of-the-art EUV imaging data obtained at 0.38~AU, a significant fraction of kernels are unresolved, implying that previous area estimates used in modelling have often been too large, causing the inferred energy fluxes to be underestimated. Furthermore, we have shown that energy injection occurs dynamically in space and time. Therefore, the pre-injection atmosphere must be carefully considered in simulations. This highlights a need to revisit 1D flare loops models in light of these new observations from Solar Orbiter and ground-based telescopes such as the DKIST and SST. 

Typical energy fluxes that have been simulated in the past are in the range $\text{F} = 10^9-10^{11}~\text{erg}~\text{s}^{-1}~\text{cm}^{-2}$ \citep[e.g.][]{FCHROMA_2023A&A...673A.150C}. However, more extreme energy fluxes up to $5\times10^{11} ~\text{erg}~\text{s}^{-1}~\text{cm}^{-2}$ have also been modelled by \cite{Kowalski+2017ApJ...836...12K}. Some observations of large events indicate energy fluxes $>10^{12} ~\text{erg}~\text{s}^{-1}~\text{cm}^{-2}$, for example, those using high-resolution optical observations of flare ribbons \citep[e.g.][]{krucker_2011ApJ...739...96K}. High energy flux estimates may become more and more common with the enhanced spatial resolution of current instrumentation. With such high flux densities, various other transport effects, such as return current losses and beam instabilities, must be taken into account \citep[e.g.][]{Alaoui+2024ApJ...974..177A}. The time range used for spectral analysis to infer the beam parameters in a specific region must also be reassessed. In the past, a hard X-ray burst of duration $\sim 10-20$~s was typically considered to be from a single flare kernel and used as an injection into a single strand \citep[e.g.][]{Kerr+2022FrASS...960856K}. In some cases, heating durations up to a minute were even simulated \citep[e.g.][]{Reep_2017}. The observations presented here indicate that this may not be a realistic heating time for a single localised brightening. The energy injection times should reflect the latest results that provide an upper bound of a few seconds. Furthermore, obtaining accurate electron beam parameters is often complicated by uncertainties in the low-energy cut-off value and electron trapping in the corona. Therefore, the implications of the results presented here on 1D radiation hydrodynamic flare loop models warrant further in-depth study.

\section{Conclusions}
In this work, data from the 2024 major flare campaign of Solar Orbiter were analysed. Those data showed that during the impulsive phase of the M2.5 GOES-class flare, SOL2024-03-23T23:41, when electron acceleration was strong, a significant fraction (approximately 50\%) of the EUV kernels measured by \HRIEUV in the 174~\r{A} passband were unresolved at a pixel scale of 135~km/pix and 2-pixel FWHM of the PSF core. These results are consistent with recent observations of chromospheric blobs in flare ribbons at high spatial resolution with SST and the DKIST \citep{Thoen_Faber+2025A&A...693A...8T, Yadav+2025}. In those works, chromospheric blobs with FWHM sizes in the range 140--200~km and 320--455~km were identified by SST and the DKIST, respectively. This corresponds to between 1--4 \HRIEUV pixels at 0.38~AU. Structures of this size are not resolvable by \HRIEUVNOSPACE. Our results therefore agree with those obtained by SST and the DKIST, depsite the fact that \HRIEUV probes a different layer of the solar atmosphere than the SST and the DKIST. In addition, we showed that for the flare studied, energy deposition occurred in a new region along the flare arcade in quick succession (on timescales of several seconds). Furthermore, an average EUV kernel time profile was derived. An upper bound was placed on the average half-maximum heating time of only a couple of seconds in duration, with correspondingly fast cooling times. Our best fit of the average EUV kernel time profile gave $\text{t}_{\text{half-max, rise}} = 1.7\pm0.3~\text{s}$  and $\text{t}_{\text{half-max, decay}} = 2.3^{+0.7}_{-0.4}~\text{s}$. Together, these results highlight the transient and fragmented nature of solar flare energy release on small spatial scales. The findings presented here have important implications on our understanding of flare energy deposition. The full extent to which these observations alter our understanding warrants detailed modelling investigation. However, what remains strikingly clear is that the energy release process in flares is highly fragmented in space and time. Finally, we conclude that the bulk energy release process of solar flares consists of many smaller scale ($\lesssim 1~\text{Mm}^2$) events  that behave like multiple flare-like energy release episodes occurring in quick succession.

\begin{acknowledgements}
We thank the anonymous referee for their constructive feedback and insightful comments. We are also grateful to Tom Van Doorsselaere for his contribution to the interpretation of the results presented here. Solar Orbiter is a space mission of international collaboration between ESA and NASA, operated by ESA. The EUI instrument was built by CSL, IAS, MPS, MSSL/UCL, PMOD/WRC, ROB, LCF/IO with funding from the Belgian Federal Science Policy Office (BELSPO/PRODEX PEA 4000112292 and 4000134088); the Centre National d’Etudes Spatiales (CNES); the UK Space Agency (UKSA); the Bundesministerium für Wirtschaft und Energie (BMWi) through the Deutsches Zentrum für Luft- und Raumfahrt (DLR); and the Swiss Space Office (SSO). HC and LAH had additional support from an EUI guest investigator grant issued by the EUI instrument team. The STIX instrument is an international collaboration between Switzerland, Poland, France, Czech Republic, Germany, Austria, Ireland, and Italy. LAH is supported by a Royal Society-Research Ireland University Research Fellowship (URF$\backslash$R1$\backslash$241775).

\end{acknowledgements}

\bibliographystyle{aa}
\bibliography{references}

\begin{appendix}

\section{A point source observed by an instrument with a PSF of 2-pixels at FWHM}\label{appendix:2pix_psf}
\begin{figure*}
    \centering
    \includegraphics[width=\textwidth]{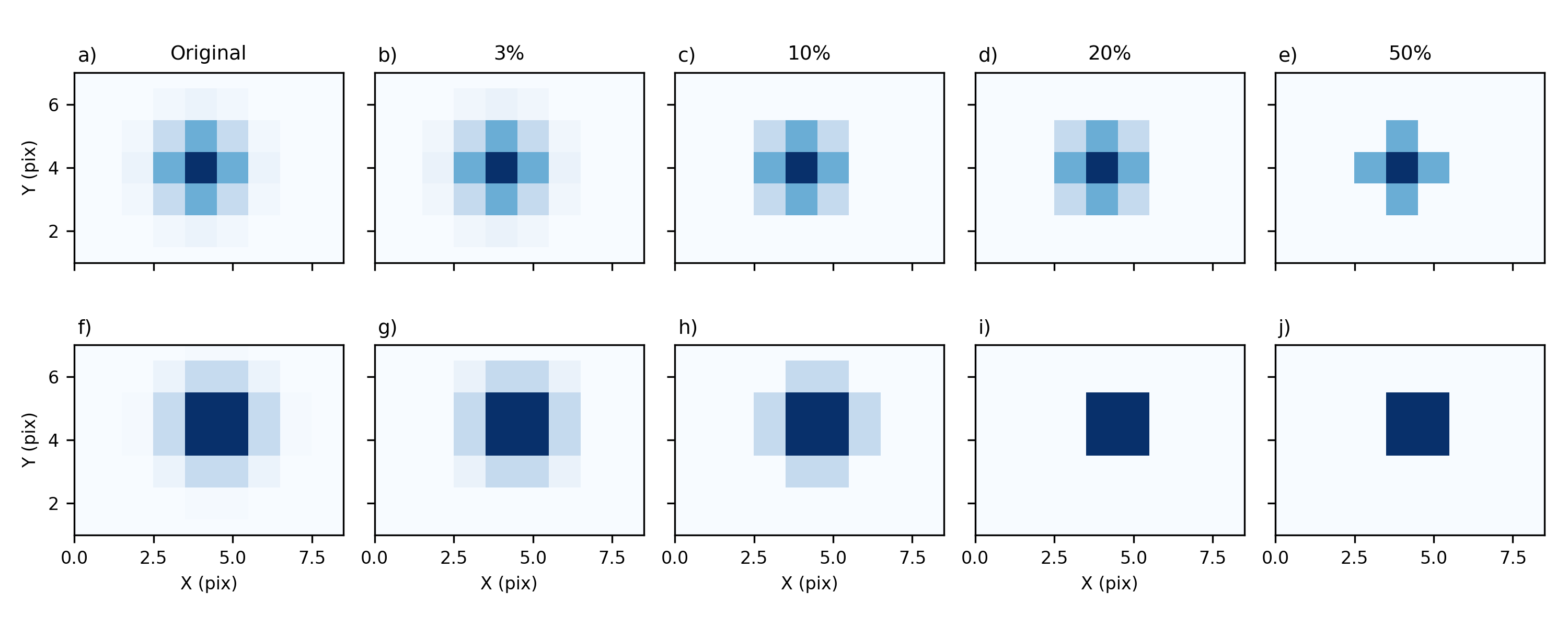}
    \caption{The effect of a 2-pixel PSF FWHM. The first column, first row, shows a point source located at the centre of a pixel smeared out due to a 2-pixel PSF. The second rows shows the same for a point source centred at exactly the intersection of four pixels. The remainder of the columns demonstrate the effect of applying a threshold on the source in both cases.}
    \label{fig:point_source_example}
\end{figure*}

In Figure \ref{fig:point_source_example} we demonstrate the effect of a 2-pixel PSF on a point source located at two extreme positions --- at the centre of a pixel and at the intersection of four pixels (top and bottom row, respectively). The effect of applying a threshold is demonstrated in the final four columns. Naturally, as the threshold increases the number of pixels with non-zero counts reduces, decreasing the size of the kernel that would be identified from processed data. In this work, a fixed cut-off threshold is used for the entire time series. As a result, fainter kernels are more affected by artificial shrinkage. We therefore caution the reader not to over-interpret the shape of the size distributions shown in Figure \ref{fig:kernel_areas_histogram} in the low-intensity range. This highlights the fact that point sources that have been artificially shrunk due to data preprocessing may appear as kernels with up to 9 pixels. Further, kernels with between 10 and 20 pixels are likely unresolved by \HRIEUVNOSPACE.

\section{Kernel size histograms at $50\%$ cutoff} \label{appendix:kernel_size_50pc}

\begin{figure*}
    \centering
    \includegraphics[width=\textwidth]{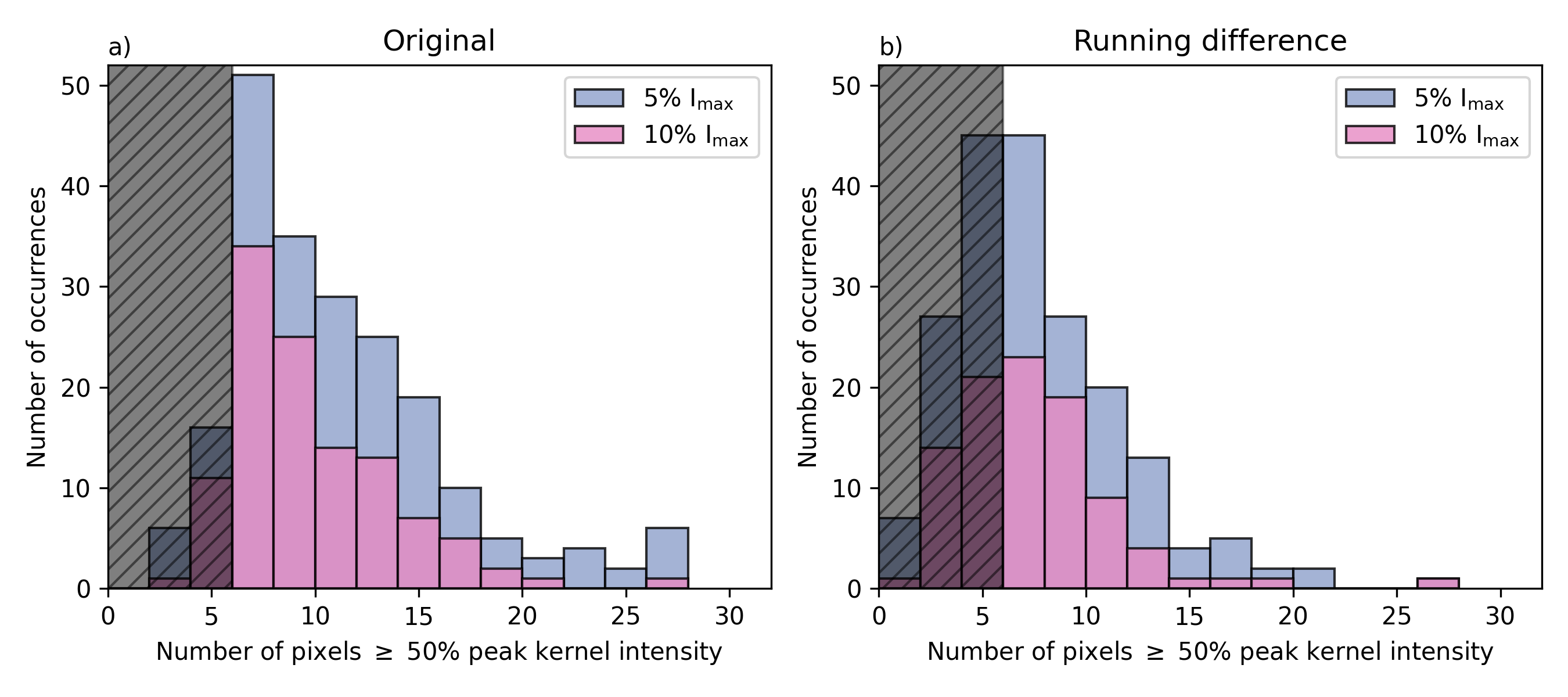}
    \caption{Kernel sizes distributions where the size is defined as the number of pixels that have $\geq 50\%~\text{I}_{\text{kernel, peak}}$. The shaded area masks kernels that contain $< 6~\text{pixels}$. This corresponds to the size of an unresolved point source clipped at 50\% of its peak intensity as shown in Figure \ref{fig:point_source_example}.}
    \label{fig:kernel_sizes_above_50p}
\end{figure*}

Figure \ref{fig:kernel_sizes_above_50p} shows the frequency distribution of kernel sizes defined as the number of pixels within a kernel that have $\geq 50\%~\text{I}_{\text{kernel, peak}}$. Kernels whose watershed label did not extend beyond the 50\% were flagged, in order to remove any bias introduced by using a fixed threshold. The distributions now have turnovers below 6 pixels because faint kernels whose boundaries were clipped by the fixed threshold have been filtered out of the dataset. Approximately 40\% of kernels contain less than 6 pixels above the 50\% peak kernel intensity level when the 5\% threshold was applied to the data prior to segmentation on the running difference sequence. For comparison, a point source smeared out by an instrumental PSF with a 2-pixel FWHM contains 5 pixels above 50\% of the peak kernel intensity when it falls on the centre of a pixel (see panel e) of Figure \ref{fig:point_source_example}).

\section{Average EUV kernel time profiles}

\begin{figure*}
    \centering
    \includegraphics[width=\textwidth]{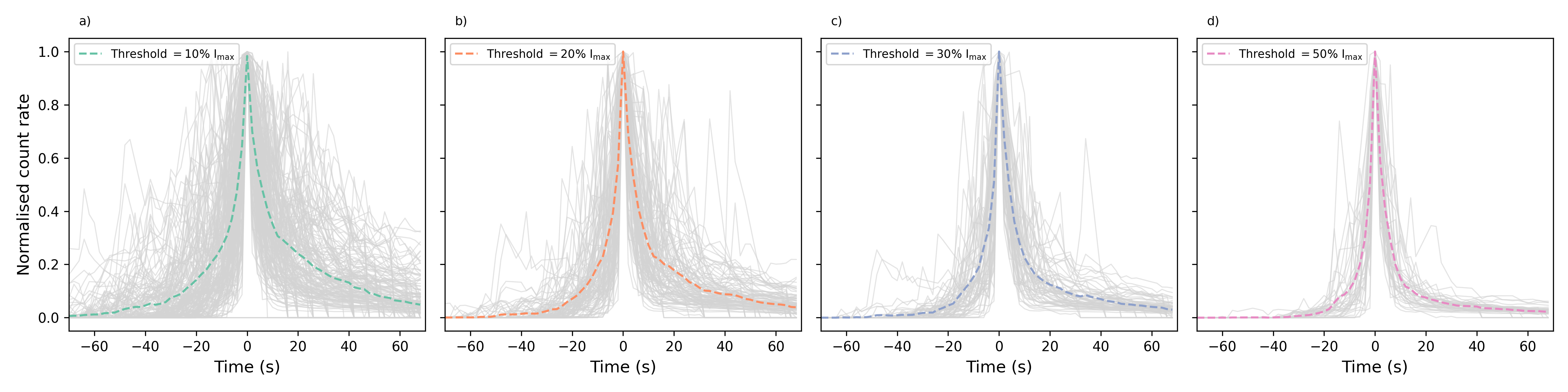}
    \caption{Average time profiles of EUV kernels given different kernel identification thresholds. Beneath, individual kernel light curves are shown in light grey. There is a significant amount of temporal variation around the main peak in many profiles. This indicates that there is fine structure in the energy release process and the average profile provides an approximate upper bound on the overall duration of a single burst.}
    \label{fig:average_lc_w_bkg}
\end{figure*}

The average EUV kernel time profiles for several thresholds were calculated and are shown in Figure \ref{fig:average_lc_w_bkg}. They have a very sharp peak. Initially, there is a gradual onset before a sharp, impulsive rise to the peak. The time profiles are asymmetric, with slightly longer decay times. The rise time for the signal to increase from 50\% to 100\% of the peak intensity in the kernel is $\text{t}_{\text{half-max, rise}}= 3.6, 2.1, 2.0$~s when the $10\%, 30\%$ and $50\%$ thresholds were applied in the pre-processing step, respectively. The time for the signal to decay back to $50\%$ of $\text{I}_{\text{max}}$ for the $10\%, 30\%$ and $50\%$ thresholds is $\text{t}_{\text{half-max, decay}} = 5.6, 4.0, 3.0$~s, respectively. This result highlights the incredibly transient nature of flare energy deposition. Since we are undersampling the profile temporally (at 2~s cadence), we often miss the peak of the profile. There are also multiple peaks in each kernel time profile at this spatial resolution.  As a result, we conclude that the impulsive flare energy injection time in a single reconnection event is less a few seconds in duration.

\end{appendix}

\end{document}